\documentstyle[prl,aps,preprint]{revtex}

\begin{document}
\tighten

\title{SPIN STRUCTURE FUNCTIONS OF THE NUCLEON\thanks{Plenary
talk presented at Baryon '95, the 7th International
Conference on the Structure of Baryons, Oct. 3-7, 1995, Santa Fe, USA}
\thanks {This work is supported in part by funds provided by the
U.S.  Department of Energy (D.O.E.) under cooperative agreement
\#DF-FC02-94ER40818.}}

\author{Xiangdong Ji}

\address{Center for Theoretical Physics \\
Laboratory for Nuclear Science \\
and Department of Physics \\
Massachusetts Institute of Technology \\
Cambridge, Massachusetts 02139 \\
{~}}

\date{MIT-CTP-2480, hep-ph/9510362 {~~~}
Submitted to: {\it Baryon '95} {~~~} October 1995}

\maketitle

\begin{abstract}
I begin with a general discussion about importance
of constructing a picture of the
nucleon in terms of QCD degrees of freedom, emphasizing
the role of spin structure functions.
I then give a short overview
on the theoretical and experimental
status of the spin structure of the nucleon. Following
that, I mention several upcoming experiments
to measure the flavor and sea structure in polarized quark
distributions and the polarized gluon distribution $\Delta g(x)$.
Finally, I discuss other spin-related physics,
such as the polarizabilities of gluon fields $\chi_B $ and $\chi_E$,
the quark transversity distribution $h_1(x)$,
and the spin structure functions $G_1$ and $G_2$ at low $Q^2$.

\end{abstract}
\pacs{xxxxxx}

\section{Why Are Structure Functions Interesting?}

The answer to the question is simple: The structure
functions directly reflect the QCD degrees of freedom
manifested by quarks and gluons.
In the past three decades, our understanding of the nucleon
structure has mainly come from models: the Constituent Quark Model,
the Nambu-Jona-Lasinio Model, Bags, Strings, Hedgehogs,
Instantons, Monopoles, to name just a few. These models
are quite successful in explaining bulk properties
of the nucleon. For instance, every model is made to
fit the nucleon mass and most predicts correctly the anomalous
magnetic moment. However, these models cannot
explain more detailed aspects of the nucleon
structure such as structure functions, because no
one knows yet how to translate effective
degrees of freedoms used in the models to
quarks and gluons in QCD. [This is perhaps a bit of
pessimistic in light of the many successes of models, like the
quark models, in which a direct identification of
QCD and constituent quarks is usually made. However, the
right question to ask in QCD should be why the models
are so successful, not why they fail occasionally.]
For the same reason, the picture of the
nucleon in QCD can be very different from that in a model.
To illustrate this point, let me consider perhaps the most basic
property of the nucleon: the mass.

The mass structure of the nucleon varies dramatically
in different phenomenological models. In the constituent quark
model, the mass is a sum of the three constituent
quark masses plus a small amount of
kinetic and potential energy. In the simplest
version of the MIT bag model, the mass is a
sum of kinetic energies for three quarks plus the
vacuum energy of the bag. In QCD, a study of the
energy-momentum tensor shows that the nucleon mass
can be separated into four gauge-invariant parts\cite{jimass},
\begin{equation}
       M = M_q + M_m  + M_g + M_a \ ,
\end{equation}
where $M_q$ is the matrix element of $\int d^3 x
\psi^\dagger (-i{\bf \alpha\cdot D})\psi $ and represents
the contribution of quark kinetic and potential energies.
$M_m$ is the matrix element of $\int d^3 x \bar \psi m \psi$
and represents the contribution of quark masses.
$M_g$ is the matrix element of $\int d^3 x ({\bf E}^2 + {\bf B}^2)/2$
and represents the contribution of the gluon energy.
Finally, $M_a$ is the matrix element of
$\int d^3 (9\alpha_s/16\pi^2)({\bf E}^2-{\bf B}^2)$ and represents
the contribution of the QCD trace anomaly\cite{cdj}.

Using the deep-inelastic scattering data
on $F_2$\cite{cteq} and the $\pi-N$ $\sigma$ term and
the second-order chiral perturbation calculation for
the baryon-octet mass splitting \cite{gasser},
it was found that \cite{jimass},
\begin{eqnarray}
          M_q  &=& 270~~ {\rm MeV }, ~~~
          M_m = 160~~ {\rm MeV },   \nonumber   \\
           M_g &=& 320~~ {\rm MeV  }, ~~~
           M_a = 190~~ {\rm MeV } \ .
\end{eqnarray}
Thus the quark kinetic and potential energies contribute only
about a third of the nucleon mass. The quark masses contribute
about one-eighth. The canonical gluon energy also contributes
about a third. It was argued in Ref. \cite{jimass} that the
anomaly contribution is analogous to the vacuum energy in
the MIT bag model. Since
none of the phenomenological models gives a similar mass structure,
it is  difficult to relate the effective degrees of freedom
with the QCD quarks and gluons in a straightforward way.

\section{The Spin Structure of the Nucleon}

The holy grail in studying the spin structure of the nucleon
is to know how the spin of the nucleon is distributed among
its constituents. Intuitively, one can write done the
following decomposition of the nucleon spin,
\begin{equation}
     {1\over 2} = {1\over 2} \Delta \Sigma + \Delta g  + L_q + L_g \ ,
\end{equation}
where $\Delta \Sigma$ and
$\Delta g$ are helicities of quarks and gluons, and
$L_q$ and $L_g$ are the quark and gluon
orbital angular momenta.

In QCD, there is good news and bad news about this
decomposition. The good news is that the separation has
a field theoretical foundation, in the sense that each part
can be identified with a matrix element of a quark-gluon
operator. Indeed \cite{jaffemanohar},
\begin{eqnarray}
      \Delta \Sigma &=& \langle P+|\int d^3x \bar \psi
     \gamma^3 \gamma_5 \psi|P+\rangle\ ,  \nonumber  \\
      \Delta g &=& \langle P+|\int d^3x (E^1A^2-E^2A^1)|P+\rangle\ ,
     \nonumber \\
      L_q &=& \langle P+| \int d^3x i \psi^\dagger
    (x^1\partial^2 - x^2\partial^1)\psi
    |P + \rangle\ , \nonumber \\
     L_g &=& \langle P+|\int d^3x E_i(x^2 \partial^1
       - x^1\partial^2) A_i |P+\rangle \ .
\end{eqnarray}
The bad news is that field theory is counter-intuitive.
In fact, $\Delta g$, $L_q$ and $L_g$ are neither separately gauge
invariant nor Lorentz invariant, although their sum is.
Futhermore, because composite
operators in field theory are generally divergent, their matrix
elements are scale-dependent after renormalization. The only
component that is gauge invariant and frame-independent
is the quark helicity contribution. The frame independence is
obvious from the Lorentz structure of the matrix element. In the
literature, there have been discussions about boosting the
quark spin from rest to the infinite momentum
frame. These discussions are irrelevant to the problem at hand.

It is easy to show that in the infinite momentum frame (or light-front
coordinates) and light-like gauge ($A^+=0$) $\Delta g$
is the first moment of the gluon helicity distribution
$\Delta g(x)$, measurable in high-energy processes.
For this reason, I will talk henceforth about the spin
decomposition in the infinite momentum frame and
light-like gauge.

The renormalization scale dependence
can be calculated in perturbative
QCD. For the quark and gluon helicities, the evolution
in the leading-log approximation is the Altarelli-Parisi
equation \cite{ap},
\begin{equation}
      {d\over dt} \left(\begin{array}{c}
                      \Delta \Sigma \\
                          \Delta g
                    \end{array} \right)
    = {\alpha_s(t)\over 2\pi} \left( \begin{array}{cc}
                       0 & 0 \\
                      {3\over 2}C_F & {\beta_0\over 2}
               \end{array} \right)
        \left(\begin{array}{c}
                      \Delta \Sigma \\
                          \Delta g
                    \end{array} \right) \ ,
\label{ape}
\end{equation}
where $t = \ln {Q^2/\Lambda_{\rm QCD}^2}$, $C_F=4/3$ and $\beta_0
=11-2n_f/3$ with $n_f$ the number of quark flavors.
The solution of the equation is well-known,
\begin{eqnarray}
       \Delta \Sigma(t) & = & {\rm const} \ , \nonumber  \\
       \Delta g(t) & = & -{4\Delta \Sigma \over \beta_0} + {t\over t_0}
            \left(\Delta g_0 + {4\Delta \Sigma\over \beta_0}\right)\ .
\label{sol1}
\end{eqnarray}
Thus the gluon helicity grows logarithmically with the renormalization
scale $Q^2$. To understand
this physically, let us consider the splitting of
a helicity +1 gluon. There are four possible splitting products:
1) a quark with helicity 1/2 and an antiquark with helicity
$-1/2$; 2) a quark with helicity $-1/2$ and an antiquark with helicity
1/2; 3) a gluon with helicity +1 and another with helicity
$-1$; 4) two gluons with helicity +1. In the first two processes
there is a loss of the gluon helicity with a probability $(n_f/2)
\int^1_0 dx(x^2 + (1-x)^2)$. In the third process there is
a loss of gluon helicity with a probability $\int^1_0
dx ({x^3/(1-x)} + {(1-x)^3/ x})$. And in the last process
there is an increase of gluon helicity with a
probability $\int^1_0 dx 1/(x(1-x))$. The total helicity
change in one gluon splitting is $11/2-n_f/3>0$.  Thus the
gluon helicity increases without bound as one probes increasingly
smaller distance scales.

The evolution of the quark and gluon orbital angular momenta was first
recognized and discussed by Ratcliffe \cite{ratcliffe}. However, the
discussion is incomplete and contains a mistake.  Recently,
Tang, Hoodbhoy and I derived the following equation \cite{jth},
\begin{equation}
      {d\over dt} \left(\begin{array}{c}
                      L_q \\
                      L_g
                    \end{array} \right)
    = {\alpha_s(t)\over 2\pi} \left( \begin{array}{rr}
                       -{4\over 3}C_F & {n_f\over 3}\\
                      {4\over 3}C_F & -{n_f\over 3}
               \end{array} \right)
        \left(\begin{array}{c}
                      L_q \\
                      L_g
                    \end{array} \right) +
   {\alpha_s(t)\over 2\pi} \left( \begin{array}{rr}
                       -{2\over 3}C_F & {n_f\over 3}\\
                       -{5\over 6}C_F & -{11\over 2}
               \end{array} \right)
         \left(\begin{array}{c}
                      \Delta \Sigma \\
                          \Delta g
                    \end{array} \right) \ .
\label{hjt}
\end{equation}
 If one knows the nucleon spin composition at a perturbative
scale $Q_0^2$, one can get the spin composition at any other
perturbative scale by solving these equations. As $Q^2\rightarrow
\infty$, the solution becomes especially simple,
\begin{eqnarray}
      L_q + {1\over 2}\Delta \Sigma &=& {1\over 2}{
      3n_f\over 16 + 3n_f}\ ,  \nonumber \\
      L_g + \Delta g &=& {1\over 2}{
      16\over 16 + 3n_f}\ .
\end{eqnarray}
Thus the partition of the nucleon spin between quarks
and gluons follows the well-known partition of the nucleon momentum
\cite{gw}! If the $Q^2$ evolution is slow, then it predicts
that quarks carry only about 50\% of the
nucleon spin even at low momentum scales.

Given these theoretical comments, let us now consider the experimental
status. In the past several years, EMC/SMC and E142/143 experiments
have established conclusively that\cite{spinexp}
\begin{equation}
    \Delta \Sigma (Q^2 \sim 10 {\rm GeV}^2) \sim 0.3 \pm 0.07 \ ,
\end{equation}
that is, about 70\% of the nucleon spin is carried by $\Delta g$,
$L_q$ and $L_g$. If slow $Q^2$ variation is true,
one expects the quark orbital angular momentum also carries
about 10\% to 30\% of the nucleon spin. Using SU(3) symmetry and the hyperon
$\beta$ decay data, it was determined that \cite{ellis},
\begin{eqnarray}
     \Delta u &=& 0.83 \pm 0.03\ , \nonumber \\
     \Delta d &=& -0.43 \pm 0.03\ , \nonumber \\
     \Delta s &=& -0.10 \pm 0.03\ .
\end{eqnarray}
Thus it seems that about 10\% of the spin is carried by the strange
flavor. However, there are arguments in the literature that the
SU(3) symmetry breaking can change this number significantly
\cite{lipkin}.

What are the future opportunities in studying the spin
structure of the nucleon? First of all, we would like to find
flavor separation and sea quark polarization. An independent
determination of the flavor separation can test the
validity of SU(3) symmetry in the hyperon $\beta$-decay data.
The size of the sea quark polarization might be the key to
understand the smallness of $\Delta \Sigma$.
At present time, there are two experiments which promise to
study in detail the flavor and sea structure.
First is the HERMES experiment at
HERA \cite{hermes}. Motivated by an idea by Frankfurt et al.
and Milner and Close \cite{frankfurt},
the HERMES experiment plans to study the pion production in
\begin{equation}
      \vec{e }+ \vec{P }\rightarrow e' + \pi + X \ .
\end{equation}
Define the production asymmetry according to,
\begin{equation}
    A_{LL} = {N^{\pi^+-\pi^-}_{\uparrow}(x) - N^{\pi^+-\pi^-}_{\downarrow}(x)
     \over N^{\pi^+-\pi^-}_{\uparrow}(x) + N^{\pi^+-\pi^-}_{\downarrow}(x)} \ ,
\end{equation}
where $N^{\pi^+-\pi^-}_{\uparrow\downarrow}$ is the
number of $\pi^+$ minus the number of $\pi^-$ produced
in the current fragmentation region when the nucleon target
is polarized. It is simple to show in the simple parton model that,
\begin{equation}
       A_P = {4\Delta u^v(x) - \Delta d^v(x)\over 4u^v(x) - d^v(x)}; ~~~
      A_D = {\Delta u^v(x) - \Delta d^v(x)\over u^v(x) - d^v(x)}\  ,
\end{equation}
where $A_P$ and $A_D$ refers to asymmetries for
proton and deuteron targets, respectively.
Thus by measuring these, one can determine
the valence polarizations
of the up and down quarks separately.

The second experiment is at polarized RHIC, where one can
study polarized proton collisions\cite{rsc}.
By measuring the single spin asymmetry in
$W^{\pm}$ boson production,
\begin{equation}
   A_L^{W^+}= {\Delta u(x) \bar d(y) - \Delta \bar d(x) u(y)
        \over u(x) \bar d(y) + \bar d(x) u(y)}\ .
\end{equation}
with $u\leftrightarrow d$ for $A^{W^-}_L$, one can
extract quark and antiquark helicity distributions
independently. The Drell-Yan process with fixed targets also offers
an interesting opportunity in this direction\cite{moss}.

The polarized gluon distribution yields information on the gluon
helicity contribution to the nucleon spin. Nothing is known
about it yet experimentally. Not much is known theoretically, except
there are arguments that $\Delta g$ is probably positive.
 From QCD perturbation theory in which a positive
helicity quark is more likely to produce a positive helicity
gluon, Brodsky, Burkardt and Schmidt proposed
a polarized gluon distribution \cite{brodsky},
\begin{equation}
    \Delta g (x) = {35\over 24}[1-(1-x)^2](1-x)^4
\end{equation}
which yields $\Delta g = 0.54$. Jaffe
has recently calculated $\Delta g$ from the MIT bag model.
He argued that the positive sign is related
to $N-\Delta$ splitting\cite{jaffe}. It is peculiar though that
his result comes entirely from the bag boundary!

Experimentally, one can probe the gluon distribution
through both deep-inelastic electron scattering and
hadron-hadron scattering.
In the former process, one can learn about the gluon distribution
through $Q^2$-evolution of $g_1$ structure function, two-jet production,
or $J/\psi$ production. There experiments can be done in polarized
HERA or future ELFE machine. It seems to me that extracting
$\Delta g$ from these processes is quite difficult due to
high demanding for statistics. A better place to learn about the gluon
polarization may be at RHIC, where one can measure $\Delta g$
in a more direct way, for instance, through
jet cross section, direct photon production, or gluon
fusion processes\cite{rsc}.

\section{Other Spin-Related Physics}

In this part, I discuss three topics that are
related to the polarized nucleon.
First is the polarizabilities of the color electric
and magnetic fields when the nucleon is at its rest
frame. Second is the quark transversity distribution
in a transversely polarized nucleon.
and finally, the spin-dependent structure function $G_1$ and
$G_2$ at and near $Q^2=0$.

\bigskip
1). {\bf Polarizabilities of Color Fields}.
If a nucleon is polarized in its rest frame
with polarization vector ${\bf S}$, how do the color
fields inside of the nucleon respond? Intuitively,
due to parity conservation, the color magnetic
field orients in the same direction as the
polarization and the color electric field in the direction
perpendicular to it. In QCD, one can
define the following polarizabilities of color
fields \cite{schafer},
\begin{eqnarray}
       \langle PS|\psi^\dagger g{\bf B}\psi |PS \rangle
     = 2\chi_B M^2 {\bf S} \ ,  \nonumber \\
       \langle PS|\psi^\dagger \alpha \times
  g{\bf E}\psi|PS \rangle = 2\chi_E M^2 {\bf S} \ ,
\end{eqnarray}
How to measure $\chi$'s? This can be done in the polarized
electron scattering where the
struck quark absorbs the virtual photon and propagates
in the background color fields of the nucleon. The
effects of color fields are the final state
interactions (FSI). Here
we have a unique situation that the FSI
helps us to learn about the properties of the
nucleon, unlike in many other cases in nuclear physics
where FSI is entirely a nuisance.

In the polarized electron scattering,
one measures two spin-dependent structure functions
$G_1$ and $G_2$, defined through the
antisymmetric part of the hadron tensor,
\begin{eqnarray}
     W^A_{\mu\nu} & = &{1\over 4\pi} \int d^4\xi ~e^{i\xi\cdot q}~
  \langle PS|J_{[\mu}(\xi) J_{\nu]}(0) |PS\rangle \nonumber \\
    & = & -i\epsilon_{\mu\nu\alpha\beta}
      q^\alpha \left[ S^\beta {G_1 \over M^2}  +
       (\nu M S^\beta - (S\cdot q) P^\beta){G_2\over M^4} \right] \ ,
\end{eqnarray}
In the deep-inelastic limit, define two scaling
functions $g_1 = (\nu /M) G_1 $ and $g_2 = (\nu/M)^2 G_2$.
According to operator product expansion, we have \cite{jaffeg2,jiunrau} ,
\begin{eqnarray}
   \int^1_0 g_1(x, Q^2)dx  & = &{1\over 2}\sum_f e_f^2a_{0f}C_{0f}(\alpha_s)
    + {M^2\over 9Q^2}\sum_f e_f^2\Big[C_{2f}(\alpha_s)a_{2f} \nonumber  \\ && +
{}~ 4\tilde C_f(\alpha_s)d_{2f}
        - 4{\tilde{\tilde C_f}}(\alpha_s) f_{2f}\Big] + ... \nonumber \\
   \int^1_0 g_2(x, Q^2)x^2dx & = & {1\over 3}\sum_f e_f^2 d_{2f}\tilde
C_f(\alpha_s)
  - {1\over 3}\sum_f e_f^2 a_{0f} C_{0f}(\alpha_s)
   + {\cal O}\left(M^2\over Q^2\right) \ ,
\label{ope}
\end{eqnarray}
where $a_{0f}$ and $a_{2f}$ are the matrix elements
of the twist-two, spin-one and spin-three operators, respectively.
The index $f$ sums over quark flavors.
$C_f$'s $(1 + {\cal O}(\alpha_s))$ are coefficient
functions summarizing QCD
radiative corrections. $d_2$ and $f_2$ are the matrix elements of
some twist-three and
four operators, respectively. Knowing these two matrix
elements, one can immediately calculate the polarizabilities,
\begin{eqnarray}
           \chi_B &= {1\over 3} (4d_2 + f_2)\ , \nonumber \\
           \chi_E &= {2\over 3} (2d_2 - f_2)\ .
\label{rela}
\end{eqnarray}

Let me now quote some numbers. From the recent E143 data, it
was determined \cite{e143}
\begin{equation}
     d_2^p = 0.0054 \pm 0.005, ~~~  d_2^d = 0.004\pm 0.009\ .
\end{equation}
One the other hand, the bag model calculation yields
\cite{jiunrau,stratmann},
\begin{equation}
     d_2^p = 0.010 , ~~~ d_2^n = 0.0 \ .
\end{equation}
which is consistent with the experimental data.
The QCD sum rule calculations yield \cite{schafer2},
\begin{equation}
 d_2^p = -0.006\pm 0.003, ~~~  d_2^n = -0.017 \pm 0.005 \ .
\end{equation}
And finally a recent calculation in quenched lattice QCD gives \cite{gorkeler},
\begin{equation}
     d_2^p = -0.048\pm 0.005, ~~~ d_2^n = -0.005\pm 0.003\ .
\end{equation}
Surprisingly, both the sum rule and lattice calculations have difficult in
confronting experimental data.

\bigskip
2). {\bf Quark Transversity Distribution}. Consider a transversely-polarized
nucleon moving in the $z$ direction.
Using $q_{\downarrow}(x)$ and $q_{\uparrow}(x)$ to denote the quark
densities with polarizations $|\uparrow\downarrow\rangle
= (|+\rangle + |-\rangle)/\sqrt{2}$ where $|\pm\rangle$ are
the helicity states of quarks. Then the transversity distribution
is defined as
\begin{equation}
      h_1(x) = q_{\uparrow} (x) - q_{\downarrow}(x) \ .
\end{equation}
This distribution was first introduced by Ralston and Soper\cite{RS}
in studying polarized Drell-Yan collisions and
further studied by Artru and Mekfi\cite{artru}, Jaffe and Ji\cite{jaffeji},
and others.

$h_1(x)$ is a chiral-odd distribution, i.e. a correlation between
left- and right-handed quarks. As such, it cannot appear
in inclusive deep-inelastic scattering process. So far, two
different methods have been proposed to measure $h_1(x)$.
First is through the asymmetry in the pion
production in longitudinally polarized electron scattering
on a transversely polarized nucleon target\cite{jaffeji2}. The second is
through
the asymmetry in Drell-Yan and $Z^0$-boson production
from quark and antiquark annihilations in polarized proton-proton
scattering\cite{RS,artru,jaffeji,soffer}.

One of the most interesting aspects of $h_1(x)$ is
the sum rule. It was shown by Jaffe and Ji that $h_1(x)$
obeys the following sum rule,
\begin{equation}
      \int^1_0 dx (h_1(x) - \bar h_1(x)) = \delta q \ .
\end{equation}
where $\delta q$ is the tensor charge of the nucleon, defined
in terms of the nucleon matrix element of the tensor current.
It is easy to shown that in non-relativistic quark models, the tensor
charge is equal to the axial charge. In the MIT bag model,
it was determined that $\delta u = 1.17, \delta d = -0.29$\cite{heji}.
On the other hand, the QCD sum rule calculation
yields $\delta u = 1.00\pm 0.5$ and
$\delta d = 0.0\pm 0.5$\cite{heji}. And more recently, the calculation
in the chiral soliton model produced $\delta u=1.07$ and
$\delta d=-0.38$\cite{chiralbag}.

\bigskip

3). {\bf  $G_1$ and $G_2$ Structure Functions Near and At $Q^2=0$}.
This is mostly about CEBAF physics. Due to time (and space) limitations,
I just briefly mention
a few important topics in this direction. It is interesting to test
the Drell-Hearn-Gerasimov(DHG) sum rule derived many years ago\cite{DHG},
\begin{equation}
\int^\infty_{\nu_{\rm th}} {d\nu \over \nu}[\sigma_{3/2}-\sigma_{1/2}]
   = {2\pi^2\alpha_{\rm em} \over M^2}\kappa^2\ ,
\end{equation}
where the $\sigma_{1/2,3/2}$ refer to the inclusive photo-production
cross sections with total helicities 1/2 and 3/2, respectively,
along the photon momentum axis.
$\kappa$ is the anomalous magnetic moment of the nucleon.
Several experiments proved at CEBAF will be relevant to such test
\cite{cebaf}.
It is also interesting to measure the spin polarizability
defined as the next moment of the photon-production cross
section difference,
\begin{equation}
    \gamma = -{1\over 4\pi^2}\int^\infty_{\nu_{\rm th}} {d\nu \over \nu^3}
    [\sigma_{3/2}-\sigma_{1/2}] \ ,
\end{equation}
and to compare it with chiral perturbation calculation\cite{holstein}.
A third topic is to study $Q^2$ dependence of $G_1$ sum rule.
Defining,
\begin{equation}
      \Gamma(Q^2) = {Q^2\over 2M^2}\int^{\infty}_{Q^2/2}
       G_1(\nu, Q^2) {d\nu\over \nu} \ ,
\end{equation}
one can expand at low-$Q^2$\cite{jiunrau},
\begin{equation}
      \Gamma(Q^2) = 1.396 - 8.631 Q^2 + \alpha Q^4 + ... \ .
\end{equation}
The first two terms are known from elastic scattering properties of the
nucleon and the DHG sum rule. It is possible to
calculate the coefficient $\alpha$ in low-energy theories, like
chiral-perturbation theory. Again, one
can test the calculation by measuring the  $Q^2$ dependence of the
generalized DHG sum rule.
Finally, I would like to mention the hadronic contribution
to hyperfine splitting of hydrogen atom, which is a
complicated integral of $G_1$ and $G_2$ structure functions
at low $Q^2$\cite{hughes}. Recently, P. Unrau has made an estimate
of the contribution and found it at the level of 0.5 ppm \cite{unrau}.
With better knowledge of $G_1$ and $G_2$, one hopes to compute
the contribution at a better precision.

\section{Summary and Conclusion}

I believe the goal of this field is to study and eventually
understand the structure of hadrons in terms of
QCD degrees of freedom: quarks and
gluons. In this regard, much progress has been made in the
last few years.
Experimentally, we have tested the Bjorken sum
rule at the ten percent level, an important check on our
understanding of experiments and QCD analysis. We have measured
with good precision the quark helicity contribution to the
nucleon spin. Theoretically, we now know how to generalize
operator production expansion to any hard scattering process,
to classify quark and gluon distribution
functions, and to calculate perturbative
corrections at the first few orders. However,
much lies ahead of us. We would like to understand better the flavor
and sea separation of quark helicity distributions. We
would like to know the polarized gluon distribution.
We need to measure the higher twist effects to better
accuracy. We shall study  systematically at CEBAF
$G_1$ and $G_2$ structure functions at low $Q^2$. Finally, there is
a lot to learn from hadron final states. Thus, I
conclude that in the area of spin physics these are exciting times.

\acknowledgements
I would like to thank the organizers of the spin session
for this wonderful opportunity to talk about spin physics.

\end{document}